\documentstyle[sprocl]{article}

\def\go{\mathrel{\raise.3ex\hbox{$>$}\mkern-14mu
             \lower0.6ex\hbox{$\sim$}}}

\begin{document}
\title{DYNAMICS OF PSR J0045-7319/B-STAR BINARY AND 
NEUTRON STAR FORMATION}
\author{ DONG LAI }
\address{Theoretical Astrophysics, 130-33, California Institute of
Technology\\ Pasadena, CA 91125}

\maketitle\abstracts{
Recent timing observations have revealed the presence
of orbital precession due to spin-orbit coupling and rapid orbital decay
due to dynamical tidal interaction in the PSR J0045-7319/B-star
binary system. They can be used to put concrete constraints
on the age, initial spin and velocity of the neutron star. 
}

{\it To be published in 
{\rm The Proceedings of the 18th Texas Symposium on Relativistic
Astrophysics} (Chicago, IL, 12, 1996) 
Editors: A. Olinto, J. Frieman, and D. Schramm
(World Scientific Press)}

\section*{Introduction}

\noindent
One of the fundamental questions in the studies of pulsars concerns
the physical conditions of neutron star at birth. 
Of particular interest is the initial spin periods
and velocities of pulsars, as they are related to such issues
as supernova explosion mechanism and gravitational wave emission
from core collapse. 
The PSR J0045-7319 binary (containing a $0.93$\,s radio pulsar
and a massive B-star companion in an eccentric, 51 days
orbit~\cite{Kaspi94}) is unique and important 
in that it is one of the two binary pulsars discovered so far that
have massive main-sequence star companions (The other one is 
PSR B1259-63). These systems evolve from MS-MS binaries when one of
the stars explode in a supernova to form a neutron star. 
Thus the characteristics of such pulsar binaries can potentially
be used to infer the physical conditions of neutron star
formation. The PSR J0045-7319 system, in particular, 
owing to its relatively small orbit and ``clean'' environment
(the mass loss from the B-star is negligible), exhibits
interesting dynamical orbital behaviors,
which allow for concrete constraints on the initial
spin and kick of the pulsar. 

\section*{Dynamics of PSR J0045-7319/B-star Binary}

\hskip 0.6truecm {\bf Spin-Orbit Coupling:}
Based on earlier timing data, it was suggested
that classical spin-orbit coupling is observable
in the PSR J0045-7319/B-star binary~\cite{Lai95}.
This spin-orbit coupling results from the flattening of the B-star 
due to its rapid rotation: the dimensionless distortion is related to
the spin rate $\Omega_s$ by $\varepsilon\sim\Omega_s^2/(GM_c/R_c^3)$,
and the resulting quadrupole moment is $Q\sim kM_cR_c^2\varepsilon$,
where $M_c,\,R_c$ are the mass and radius of the B-star, $k$ is
a constant measuring the mass concentration inside the star.
There are two effects associated with this spin-induced quadrupole
moment: (i) The {\it advance of periastron} 
due to perturbation in the interaction potential $\Delta V\sim
GM_pQ/r^3$ (where $M_p$ is the pulsar mass, $r$ is the separation);
(ii) When the B-star's spin ${\bf S}$
is misaligned with the orbital angular momentum ${\bf L}$,
there is an interaction torque $N\sim (GM_pQ/r^3)\sin\theta$
(where $\theta$ is the angle between ${\bf S}$ and ${\bf L}$)
between the spin and the orbital motion, giving rise to
{\it precessions of ${\bf S}$ and ${\bf L}$} around a fixed 
${\bf J}={\bf L}+{\bf S}$. Both of these effects have been
confirmed by recent observation~\cite{Kaspi96}.

{\bf Rapid Orbital Decay:} 
Recent timing data also reveal that the orbit is 
decaying~\cite{Kaspi96}, on a timescale of 
$P_{\rm orb}/\dot P_{\rm orb}=-0.5$ Myr (shorter than the
lifetime of the $8.8M_\odot$ B-star and the characteristic age 
of the pulsar). Since mass loss from the B-star is negligible
(as inferred from dispersion measure variation),
the orbital decay must have a dynamical origin.
It was suggested that 
dynamical tidal interaction can do job~\cite{Lai96}: Each time the
pulsar passes close to the B-star, it excites internal oscillations
(mainly g-modes) in the star, transferring orbital energy to
the stellar oscillations. An interesting prediction of this theory
is that in order for the energy transfer to be sufficiently large
to explain the observed orbital decay rate,
${\bf S}$ and ${\bf L}$ must be not only misaligned
but also more or less anti-aligned. 
The reason that {\it retrograde rotation} can significantly
increase the tidal strength is the following:
During a periastron passage, the most strongly
excited modes are those (i) propagating in the same direction
as the orbital motion, (ii) having frequencies in the
inertial frame comparable to the ``driving frequency'' (equal
to twice of orbital frequency at periastron), and (iii) coupling 
strongly to the tidal potential. 
Since the higher-order (lower frequency) g-modes have smaller
couping coefficients than the low-order ones, the trade-off
between (ii) and (iii) implies that the dominant
modes in energy transfer are those with frequencies higher than the
resonant mode. If the B-star were nonrotating, the dominant modes
would be g$_5$-g$_9$, and the inferred $\dot P_{\rm orb}$ would be 
two orders of magnitude too small to explain the observed value.
However, a retrograde rotation ``drags'' the
wave modes backwards and reduces the mode frequencies 
in the inertial frame. As a result, energy
transfer is dominated by lower-order modes, which couple much more
strongly to the tidal potential. At $\Omega_s=-0.4(GM/R^3)^{1/2}$
(projected along the ${\bf L}$ axis), for example, the dominant modes
are g$_3$-g$_5$, and the energy transfer increases by two orders
of magnitude as compared to the nonrotating value. 
A recent analysis~\cite{Kumar} of the radiative damping of g-modes
indicates that an additional ingredient, i.e., differential rotation,
is needed to compensate for the longer damping times of lower-order
g-modes. 

\section*{Constraints on the Initial Conditions of PSR J0045-7319}

\hskip 0.6truecm {\bf Evidence for Supernova Kick:} 
As mentioned before, the PSR/B-star binary evolves from a
MS-MS binary. At this earlier stage, the B-star's spin
is most likely to be aligned with the orbital angular momentum.
The only way to transform this aligned configuration into the
current misaligned configuration is that the supernova was asymmetric
and gave the pulsar a kick~\cite{Lai95,Kaspi96}, 
and the kick velocity must have nonzero
components (i) in the direction out of the original orbital plane,
and (ii) in the direction opposite to and with magnitude
larger than the original orbital velocity. 
Let the total mass, semimajor axis, eccentricity
of the system before and after the supernova be $(M_i,\,a_i,\,0)$ and
$(M_f,\,a_f,\,e_f)$. The kick velocity is given by
\begin{equation}
|V| = (GM_f/a_f)^{1/2}\left[2\xi-1+\xi\eta^{-1}
-2(1-e_f^2)^{1/2}\xi^{3/2}\eta^{-1/2}\cos\theta\right],
\end{equation}
where $\eta=M_f/M_i<1$ and $\xi=a_f/a_i$. With $(1+e_f)^{-1}\le
\xi\le (1-e_f)^{-1}$ and $125^o\le\theta\le 155^o$ (as constrained
by the measurement of precessions~\cite{Kaspi96} and for retrogarde
rotation; tighter constraint can be obtained using the observed
surface velocity of the B-star, but it depends on 
the precession phase), we get $|V|\go (GM_f/a_f)^{1/2}
\simeq 125$ km\,s$^{-1}$, where we have used the current observed 
values for $a_f$ and $e_f$. Orbital evolution since the
supernova tends to make $a_f$ and $e_f$ larger, hence 
decreases this lower limit.

{\bf Age of the Binary and Initial Spin of the Pulsar:}
The theory of dynamical-tide induced orbital decay gives a scaling 
relation~\cite{Lai96}
\begin{equation}
\dot P_{\rm orb}\propto P_{\rm orb}^{-7/3-4\nu}(1-e)^{-6(1+\nu)},
\end{equation}
and a similar relation for $\dot e$, 
where $\nu$ lies in the range $0.2-1.0$, reflecting 
the uncertainty in the rotation rate. The proportional 
constant depends on the (uncertain) mode damping time. 
Using the observed value of $\dot P_{\rm orb}$ for the current
system, this constant can be fixed, and the equations can be 
integrated backward in time. It was found that regardless of
the uncertainties, the age of the binary since
the supernova is less than $1.4$ Myr. This is significantly
smaller than the characteristic age ($3$ Myr) of the pulsar, 
implying that the latter is not a good age indicator.
The most likely explanation for this discrepancy is that
the initial spin period of the pulsar is close to its
current value. Thus the pulsar was either formed rotating very 
slowly, or has suffered spin-down due to accretion in the first $\sim
10^4$ years (the Kelvin-Holmholtz time of the B-star) after the
supernova (E.~van den Heuvel, private communication).

\section*{Acknowledgments} 
This research is supported by the Richard
C. Tolman Fellowship at Caltech and NASA Grant NAG 5-2756.

\end{document}